\documentclass[
    ,final            
  ]
  {aipproc}

\layoutstyle{8x11double}

\begin{document}

\title{Theory of the electronic structure of dilute bismide and bismide-nitride alloys of GaAs: Tight-binding and k.p models}


\author{Muhammad Usman}{
  address={Tyndall National Institute, Lee Maltings, Dyke Parade, Cork, Ireland.}
}

\author{Christopher A. Broderick}{
  address={Tyndall National Institute, Lee Maltings, Dyke Parade, Cork, Ireland.}
  ,altaddress={Department of Physics, University College Cork, Cork, Ireland.}
}

\author{Eoin P. O'Reilly}{
  address={Tyndall National Institute, Lee Maltings, Dyke Parade, Cork, Ireland.}
  ,altaddress={Department of Physics, University College Cork, Cork, Ireland.} 
}

\begin{abstract}
The addition of dilute concentrations of bismuth (Bi) into GaAs to form GaBi$_x$As$_{1-x}$ alloys results in a large reduction of the band gap energy (E$_{g}$) accompanied by a significant increase of the spin-orbit-splitting energy ($\bigtriangleup_{SO}$), leading to an E$_{g} < \bigtriangleup_{SO}$ regime for $x \sim 10$\% which is technologically relevant for the design of highly efficient photonic devices. The quaternary alloy GaBi$_x$N$_y$As$_{1-x-y}$ offers further flexibility for band gap tuning, because both nitrogen and bismuth can independently induce band gap reduction. This work reports sp$^3$s$^*$ tight binding and 14-band \textbf{k}$\cdot$\textbf{p} models for the study of the electronic structure of GaBi$_x$As$_{1-x}$ and GaBi$_x$N$_y$As$_{1-x-y}$ alloys. Our results are in good agreement with the available experimental data. 
\end{abstract}

\maketitle


\section{Introduction}

The highly mismatched semiconductor alloy GaBi$_{x}$As$_{1-x}$, comprised of dilute concentrations of bismuth (Bi) incorporated in GaAs, is an attractive candidate for the design of highly efficient mid- and far-infrared optical \cite{Sweeney_ICTON_2011} and spintronic \cite{Fluegel_PRL_2006} devices. It has been shown, both experimentally \cite{Batool_JAP_2012} and theoretically \cite{Usman_PRB_2011}, that isovalent substitution of a small fraction of Bi atoms in GaAs strongly reduces the band gap energy (E$_g$) by $\approx$90 meV/\% Bi replacing As. A giant bowing of the spin-orbit-splitting energy ($\bigtriangleup_{SO}$) has also been observed in GaBi$_{x}$As$_{1-x}$, leading to the onset of an E$_{g} < \bigtriangleup_{SO}$ regime. This characteristic is of fundamental importance as it opens the possibility to suppress the dominant CHSH Auger recombination losses suffered by conventional III-V telecomm lasers operating at high temperatures~\cite{Sweeney_ICTON_2011}. Therefore, an understanding of the electronic structure of GaBi$_x$As$_{1-x}$ is crucial both from a fundamental perspective, as well as for potential device applications. 

Co-alloying of Bi and N in GaAs further opens up the possibility of precise strain control in GaBi$_x$N$_y$As$_{1-x-y}$ thereby offering a large and flexible parameter space for band structure engineering. This work analyses the combined impact of Bi and of N on the electronic structure of GaBi$_x$N$_y$As$_{1-x-y}$, showing that N predominantly affects the conduction band and Bi the valence band, with both effects largely being independent of each other. A detailed chart of the variations in the E$_{g}$ and E$_g$-$\bigtriangleup_{SO}$ energies as a function of Bi and N composition is plotted as a guide for material and device design and analysis.
 
\section{Methodologies}

We have developed a nearest neighbor sp$^3$s$^*$ tight binding (TB) Hamiltonian to investigate the electronic structure of the dilute bismide and the bismide-nitride alloys of GaAs, including the effects of atomic relaxation determined using a valence force field model~\cite{Usman_PRB_2011}. By performing large, ordered GaBi$_x$Y$_{1-x}$ (Y=As,P) supercell calculations using our models, we demonstrate that isovalent Bi substitution introduces Bi-related defect states which interact with the host GaY (Y=As,P) matrix valence band edge via a Bi composition dependent band anti-crossing (BAC) interaction. The band anti-crossing parameters are then introduced into 12- and 14-band \textbf{k}$\cdot$\textbf{p} models for GaBi$_x$As$_{1-x}$ and for GaBi$_{x}$N$_{y}$As$_{1-x-y}$ to investigate the alloy electronic structure. The details of the derivation procedures are reported elsewhere \cite{Broderick_12band_2012}.   

\section{Results and Discussions}

Figures 1 (a), (b), and (c) compare band dispersions computed from the tight binding (TB) and the \textbf{k}$\cdot$\textbf{p} models in the vicinity of the $\Gamma$-point for Ga$_{32}$As$_{32}$, Ga$_{32}$Bi$_1$As$_{31}$, and Ga$_{32}$Bi$_1$N$_1$As$_{30}$ alloy supercells. The good agreement between the TB and \textbf{k}$\cdot$\textbf{p} models confirms the presence of the band anti-crossing interaction between the impurity states and the host matrix states. 

\begin{figure*}
\includegraphics[scale=0.155]{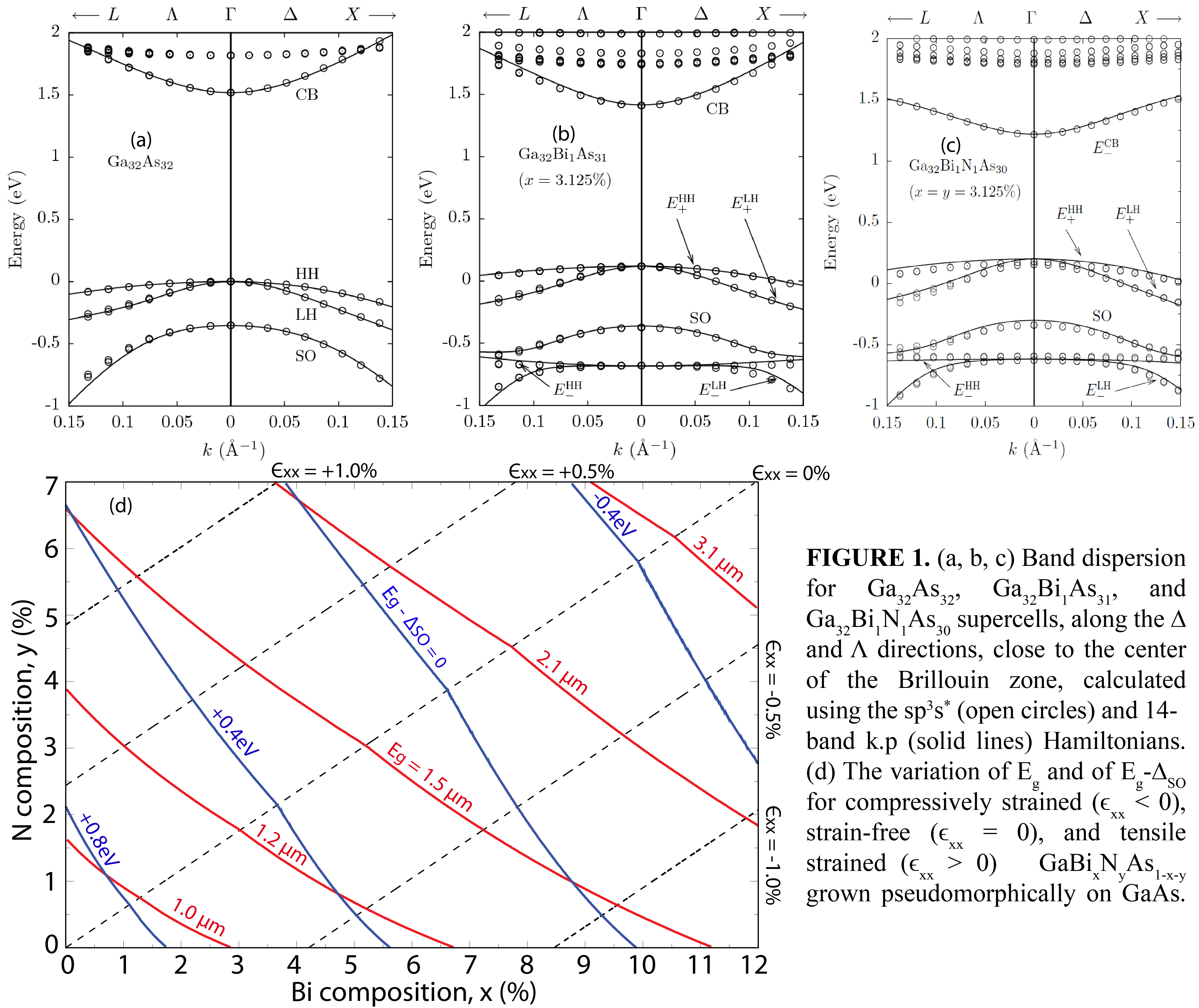}
\label{fig:Fig1}
\end{figure*}

Furthermore, by carrying out 4096-atom ordered supercell calculations, we verify that the effects of Bi and of N on the alloy electronic structure are largely independent of each other, with N-related defect states introducing a BAC interaction in the conduction band and the Bi-related impurity states introducing BAC interactions in the valence band. 

Our calculations show that the observed strong variation in E$_g$ for GaBi$_x$As$_{1-x}$ can be well explained in terms of a large upward shift in the valence band edge energy due to the BAC interaction and a significant contribution coming from a conventional alloy reduction in the conduction band edge energy \cite{Usman_PRB_2011}. The large band gap reduction with N composition in GaNAs is further enhanced by the presence of Bi in GaBi$_x$N$_y$As$_{1-x-y}$, leading to a giant E$_g$ bowing \cite{Broderick_SST_2012}. Since N only weakly perturbs the GaAs valence band structure, the bowing of $\bigtriangleup_{SO}$ in GaBi$_x$N$_y$As$_{1-x-y}$ is similar to that in the GaBi$_x$As$_{1-x}$ alloy.

Figure 1 (d) shows the variation of E$_g$ (red lines) and of E$_g$-$\bigtriangleup_{SO}$ (blue lines) for GaBi$_x$N$_y$As$_{1-x-y}$ grown pseudomorphically on GaAs with $0 \leq x \leq 12$\% and $0 \leq y \leq 7$\%, for which the layers can be under compressive strain ($\epsilon_{xx} < 0$), strain-free ($\epsilon_{xx} = 0$), or under tensile strain ($\epsilon_{xx} > 0$) as the composition is varied. A large tuning of the band gap wavelength ($\lambda$) is demonstrated, to $\lambda >$ 3.0 $\mu$m. The E$_g$-$\bigtriangleup_{SO}$ = 0 line indicates the cut-off Bi and N compositions beyond which an E$_g$ energy lower than the $\bigtriangleup_{SO}$ energy can be achieved, which is key to the suppression of the CHSH Auger loss mechanism~\cite{Sweeney_ICTON_2011}. 

In summary, our models provide a clear understanding of the electronic structure of GaBi$_x$As$_{1-x}$ and of GaBi$_x$N$_y$As$_{1-x-y}$. We conclude that these alloys have significant potential for use in highly efficient optoelectronic devices operating with reduced temperature sensitivity both at telecommunication wavelengths and in the mid-IR, critical to the development of future 'green' commercial telecommunication networks and for a wide range of sensing applications.

\begin{theacknowledgments}
This work was supported by European Union project BIANCHO (FP7-257974; MU and EOR), by Science Foundation Ireland (10/IN.1/I2994; EOR) and by the Irish Research Council under the Embark Initiative (RS/2010/2766; CAB).
\end{theacknowledgments}

\bibliographystyle{aipproc}   


\IfFileExists{\jobname.bbl}{}
 {\typeout{}
  \typeout{******************************************}
  \typeout{** Please run "bibtex \jobname" to optain}
  \typeout{** the bibliography and then re-run LaTeX}
  \typeout{** twice to fix the references!}
  \typeout{******************************************}
  \typeout{}
 }



\end{document}